# Meta-analysis in dental research

Hoi-Jeong Lim
*Department of Orthodontics, Dental Science Research Institute*
*Chonnam National University School of Dentistry*
Gwangju, South Korea
hjlim@jnu.ac.kr

*Abstract*—Recently, importance of meta-analysis is increasing in the field of dentistry, since it is not easy to settle contro versies arising from conflicting studies. Meta-analysis is the statistical method of combining results from two or more individual studies that have been done on the same topic. Merits of meta-analysis includes an increase in power, an improvement in precision, and the ability to address solution not provided by individual studies. However, it might mislead researchers when variation across studies and publication bias are not carefully taken into consideration. The purpose of this study is to help understand meta-analysis by making use of individual results in dental research paper.

## I. Introduction

The term "meta-analysis" was first used by Glass in 1976, and it began to be actively used in the medical field from the mid-1980s and in the dental field from the 2000s. Currently, the number of meta-analysis studies is increasing exponentially "Fig. 1". Meta-analysis is effectively used when there are conflicting conclusions or controversies among various studies, or when there is a need to solve problems within a short period and with a limited budget. Recently, in the field of dentistry, there has been a continuous increase in studies on similar topics, making it difficult to grasp trends through existing literature reviews, and the growing amount of related information has increased the necessity for meta-analysis.

Meta-analysis is an analysis that converts the experimental results appearing in individual studies conducted on the same or similar topics into a common effect size, allowing for objective and quantitative generalization of the experimental results. The advantages of meta-analysis include the ability to increase statistical power by combining individual studies to test hypotheses with a larger sample size, and the ability to estimate a more accurate effect size than that of individual studies by calculating the average effect size from multiple individual studies. Additionally, it can help identify the causes of conflicting research results.

However, the disadvantages of meta-analysis may include issues with attempting to combine research results of different natures that cannot be compared, similar to mixing apples and oranges. This can be resolved by excluding extreme data. There may also be issues with combining high-quality and low-quality studies without distinction, but this can be addressed by weighting the studies appropriately according to their quality levels. Furthermore, since only published studies are typically included (publication bias), there may be issues with representativeness, but this can be resolved by including unpublished studies as well[1].

The purpose of this study is to detail the procedures for conducting meta-analysis and to investigate how meta-analysis has been utilized in dental research.

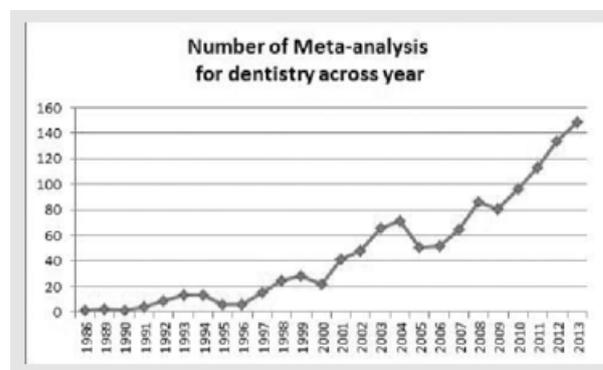

Fig. 1. Number of meta-analysis for dentistry across the year

## II. Procedure for Systematic Literature Review

### A. Setting the Hypothesis of the Research

Set up a null hypothesis or an alternative hypothesis.

Let's take dental research as an example. The purpose of a study that conducted a systematic literature review[2] was to investigate the relationship between total tooth eruption and dental caries from each published study. An example of a study that conducted both a systematic literature review and a meta-analysis is the first one[3], which aimed to identify the risk factors affecting the failure of mini-screw implants through a meta-analysis of already published controlled or uncontrolled prospective clinical trials. The second example[4] aimed to compare the failure rates of bonding and the time taken for bonding using SEP (Self-etch primers) and AE (Acid-etch technique) methods through a meta-analysis.

### B. Setting inclusion and exclusion criteria for the study, followed by literature search and paper selection.

When selecting a paper, it is necessary to provide evidence that allows for a more objective conclusion by thoroughly collecting presentations or unpublished research related to the chosen topic. The representative sources for data collection of research materials include domestic and international master's and doctoral theses, papers presented at conferences, references, books, lists of research materials, utilization of databases, and papers that are currently in the process of publication or unpublished papers. The reason for including unpublished papers is that published papers tend to favor those with significant results, making it difficult to exclude cases where effect sizes are overestimated. However, collecting all extensive data is not an easy task in terms of economics, time, and effort. In such cases, one can select an appropriate scale for analysis by randomly sampling a limited number of research papers, restricting the scope of research to a specific period by year, or narrowing down the research topic to limit the collection of prior research results.

According to Cooper (1982)[5], methods for identifying the sources of research materials include, first, an upward (ancestry) tracking approach, which involves tracing related research based on the references cited in a research presentation paper, and second, a downward (descendency) tracking research method, which utilizes computer databases such as Pubmed or Medline for information retrieval, allowing for quick searches of desired results, leading to active development in the field of meta-analysis.

Looking at the flowchart of the dental journal in "Fig. 3" with reference to "Fig. 2"[2] the number of confirmed literature through database searches was 6,911, and the number of additional literature confirmed from other information sources was 3. After removing duplicates, the remaining number of literature was 3,820, of which 3,801 were excluded.

Out of the 3,727 pieces of literature, 56 were not written in English, 15 only calculated prevalence, and 3 belonged to literature reviews, editorials, or expert opinions. Therefore, 19 original texts were obtained, of which 11 were excluded. One original text had no comparison group, 8 had unclear indices for evaluating crowding, one studied a specific group rather than a general population, and the remaining one was a study that conducted a systematic literature review. As a result, the number of studies included in the meta-analysis was 8.

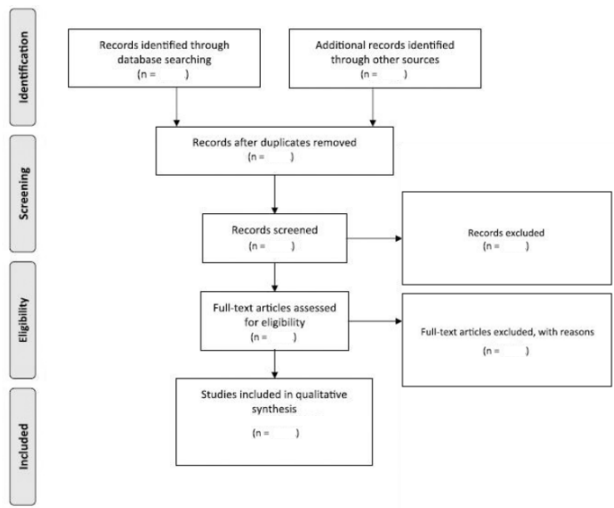

Fig. 2. PRISMA Flow Chart

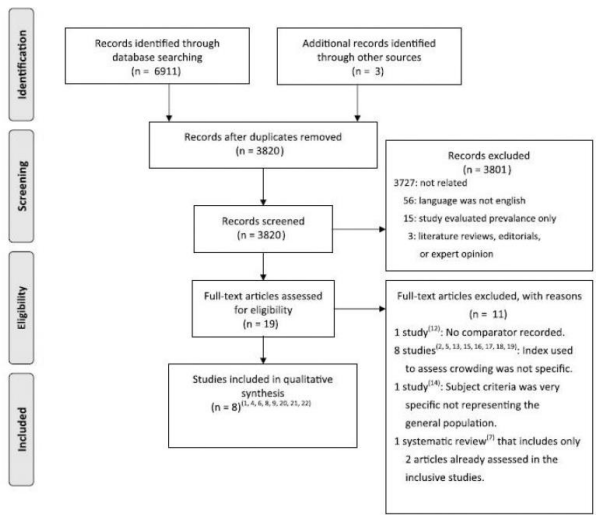

Fig. 3. PRISMA Flow Chart

TABLE I. GENERAL CHARACTERISTICS OF INCLUDED STUDIES

| Study | Age | Sex | Race | Sample size | Baseline oral health status |
|---|---|---|---|---|---|
| Hixon et al[4] | Severe crowding: not reported Excellent occlusion: mean age, 18.3 years | Severe crowding: not reported Excellent occlusion: 61% males | White | Severe crowding: 20 Excellent occlusion: 106 | Not reported |
| Roder and Arend[22] | 14-16 years | Girls | Not reported | Subgroups: not reported | Not reported |
| Katz[8] | <25 years | Not reported | White | Subjects: 160 Subgroups: not reported | Adequate condition of occlusion and teeth present |
| Addy et al[1] | 11.5-12.5 years | Role of sex is insignificant | Not reported | 2656 pairs of contralateral crowded and uncrowded teeth | Not reported |
| Helm and Petersen[20] | 33-39 years (mean age, 35.5 years) | Not reported | Not reported | Maxillary incisor crowding: 33 Mandibular incisor crowding: 39 Maxillary crowding: 41 Mandibular crowding: 51 Control group: 27 | Not reported |
| Stahl and Grabowski[21] | Children with primary dentitions (mean age, 4.5 years) and mixed dentitions (mean age, 8.9 years) | Girls: 4306 Boys: 4558 Subgroups: not reported | Not reported | Subgroups: not reported | Not reported |
| Staufer and Landmesser[5] | Age groups, 18-34 years (n = 63) and ≥35 years (n = 62) | Women: 63 Men: 62 Subgroups: not reported | Not reported | 18-34 years (n = 63) ≥35 years (n = 62) | Occlusal stability for a physiologically supported jaw and good state of care of mandibular canines and incisors |
| Alsoliman[9] | 9-13 years | Subgroups: not reported | Not reported | Not reported | Not reported |

*C. Feature extraction of research and data coding of the research*

  *1) Extraction of research characteristics.*

Generally, it is possible to extract the publication date of research papers, the age group of the research subjects, the sampling ratio between males and females, sample size, dropout of experimental subjects, whether the research subjects were sampled randomly or non-randomly, and the form of publication of the research paper (such as journal articles, books, theses, etc.). Looking at "Table 1" above[2], it can be seen that the age, gender, race, sample size, and oral health status of the research subjects have been extracted from each study.

  *2) Development of a coding manual and coding table for meta-analysis*

To avoid intra-rater error and inter-rater error, the creation of a coding manual is of utmost importance. When the coding manual is made specific and clear, it can help prevent coding errors and ultimately increase the reliability of meta-analysis research. The coding work should be conducted by one or more evaluators, and the agreement among evaluators should be calculated and reported. The coding manual below is a translation of Lipsey (2000), and you can develop your own coding manual by reviewing the items listed below. The items include information related to the paper, general characteristics of the sample, characteristics of the research design, characteristics of the dependent variable, effect size, means and standard deviations or frequencies or ratios, significance test results, and calculated effect sizes, and the coding manual can be modified according to the characteristics of the research "Fig. 4".

Fig. 4. Manual Coding

## D. Quality assessment of included studies

The quality assessment of individual studies included is an important part of a systematic literature review. The quality evaluation of each study is developed according to its specific context, leading to different assessments based on various situations. This evaluation is conducted through questions regarding the type of study, the presence of a control group, the appropriateness of outcome variable selection, the adequacy of sample size, whether measurement errors were assessed, and the appropriateness of statistical methods, among others. For example, when examining the quality assessment of studies in dental research that conducted a meta-analysis, scores were assigned to the following items, and a total score was calculated to evaluate the quality of the research.

*1) Description of the selection process: Not described (0), partially described (1), described in detail (2)*

*2) Prospective or retrospective: Retrospective study (0) or prospective study (2)*

*3) Consecutive cases: Unconsecutive (0), Consecutive (1)*

*4) Sample size (N): N<20 (0) or N≥20 (1)*

*5) Selection of outcome measure: Inappropriate (0), Partially appropriate (1), Appropriate (2)*

*6) Appropriateness of measurement error: Not measured (0), Partially appropriate (1), Appropriate (2)*

*7) Appropriateness of statistical methods: Inappropriate (0), Appropriate (1)*

Total score: Less than 7 (Low), 7-8 (Medium), 9-10 (Medium-high), 11 (High)

Eight studies received a score of less than 7 (Low), and four studies received a score of 7-8 (Medium). In conclusion, since these are not high-quality studies (High), the results should be interpreted cautiously, and the conclusions drawn from these results cannot be considered definitive.

Looking at the quality assessment of another dental research paper "Table 2"[2], it can be seen that the evaluation factors differ somewhat from those above.

- Types of research: Cross-sectional study (1), Longitudinal study (2)

- Blinding status: Not blinded (0), Blinded (2)

- Adequate reporting of selection criteria for study subjects: Describing the distribution and methodology of study subjects for each factor (1) (maximum 3 points)

- Control group: If the control group has different grades of total eruption (1), if there is a control group with normal occlusion or total eruption up to 2mm (2) (maximum 3 points)

- Validity and reliability of the method for reporting caries: Visual assessment (1), Radiographic assessment (2)

- Validity and reliability of the method for reporting total eruption: Visual assessment (1), Quantitative index (2) (maximum 2 points)

- Measurement error: Measurement error for caries detection and total eruption for each factor (1) (maximum 2 points)

- Consideration of confounding variables' effects (age, race, sex, missing teeth, and baseline oral status): Each (1)

- Subgrouping: Subgrouping of study subjects to compare age effects and the degree of total eruption, each factor (1) (maximum 2 points)

- Coding: If study subjects and variables are not coded (0), if they are coded (1)

Total score: 1-8 (Low), 9-16 (Moderate), 17-24 (High)

One study received a score of 1 to 8 (Low), and seven studies received a score of 9 to 16 (Moderate). Similarly, since these are not high-quality studies, the results should be interpreted with caution.

TABLE II.    QUALITY ASSESSMENT OF INCLUDED STUDIES

| Study | Study type[a] | Blinding[b] | Adequate reporting[c] | Comparator[d] | Validity, reliability of caries recording method[e] | Validity, reliability of crowding recording method[f] | Error of measurement[g] | Confounding factors[h] | Subgrouping[i] | Coding[j] | Score/grade |
|---|---|---|---|---|---|---|---|---|---|---|---|
| Hixon et al[4] | 1 | - | 2 | 3 | 3 | 1 | 2 | 3 race, missing teeth, sex; NS | 1 crowding | - | 16/moderate |
| Roder and Arend[22] | 1 | - | 2 | 2 | 1 | 2 | - | 2 missing teeth, sex; NS | - | - | 10/moderate |
| Katz[8] | 1 | - | 2 | 1 | 3 | 2 | - | 4 race, sex, missing teeth, baseline oral status | 1 crowding | - | 14/moderate |
| Addy et al[1] | 1 | - | 2 | 2 | 1 | 1 | - | 4 race, sex, missing teeth, baseline oral status; NS | - | - | 11/moderate |
| Helm and Petersen[20] | 1 | 2 | 3 | 2 | 1 | 1 | - | 3 age, sex, missing teeth | - | - | 13/moderate |
| Stahl and Grabowski[21] | 1 | - | 2 | 2 | 1 | 1 | - | 1 missing teeth | - | - | 8/low |
| Staufer and Landmesser[6] | 1 | - | 3 | 3 | 1 | 2 | 1 crowding | 3 age, missing teeth, baseline oral status | 2 | - | 16/moderate |
| Alsoliman[9] | 1 | - | 2 | 2 | 1 | 1 | 1 caries | 2 missing teeth, sex; NS | - | 1 | 11/moderate |

NS, Not significant.
[a]Longitudinal study: 2 points; cross-sectional study: 1 point (maximum 2 points); [b]Blinding: 2 points; [c]Adequate reporting on subject criteria, subject distribution, and methodology: 1 point for each factor (maximum 3 points); [d]Control group of normal occlusion or minimal crowding (up to 2 mm): 2 points; control of different grades of crowding: 1 point (maximum 3 points); [e]Radiographic assessment: 2 points; visual examination: 1 point (maximum 2 points); [f]Quantitative index: 2 points; visual: 1 point (maximum 2 points); [g]Error of measurement for caries detection and measurement of crowding: 1 point for each factor (maximum 2 points); [h]Effect of confounding factors considered (age, race, sex, missing teeth, and baseline oral status): 1 point for each factor. When there was no sex-related difference between caries incidence, 1 point was given because no correction was needed. When confounding factors had no effect (NS): 1 point was given for each factor (maximum 5 points); [i]Subgrouping subjects to compare the effect of age and severity of crowding: 1 point for each factor (maximum 2 points); [j]Coding of subjects and variables: 1 point.

## III. PROCEDURE FOR CONDUCTING A META-ANALYSIS

The detailed procedures for conducting a meta-analysis using the coded data after performing a systematic literature review are as follows.

A. The following review checks whether the basic premises required for meta-analysis are met. There must be a sufficient number of prior studies on the topic to be integrated, and the research design of the prior studies used in the meta-analysis must be experimental research that includes both a control group and an experimental group. Additionally, the studies should report the means and standard deviations of each group, sample sizes, and significance levels. Even if the above information is not provided, the studies should present statistical values such as t-tests, F-tests, or correlation coefficients (r)[8].

B. The data for the outcome variable can be classified into binary data, continuous data, ordinal data, and survival data, and by reviewing the types of summary statistics related to these, we determine in advance which summary statistics, that is, effect sizes, will be calculated for the outcomes. Effect sizes may include standardized mean difference (SMD), odds ratio (OR), risk ratio (RR), correlation coefficient (r), ratios, p-values, etc. Here, we will only compare and review continuous data and binary data; the continuous outcome variable is evaluated as mean difference, while the outcome variable for binary data is expressed as a ratio. Looking at the dental example, the standardized mean difference (SMD) was calculated as the effect size (Effect Size: ES) by extracting the mean and standard deviation of anchorage loss for both the mini-screw implant (MI) group and the previously used conventional anchorage group (CA), along with the sample size, as shown in the formula below.

$$ES_{sm} = \frac{\overline{X_e} - \overline{X_c}}{S_{pooled}}$$

$$\text{where } S_{pooled} = \sqrt{\frac{(n_1 - 1)S_1^2 + (n_1 - 1)S_2^2}{n_1 + n_2 - 2}}$$

In another dental example, we examined the adhesive failure rates between the self-etch primer (SEP) group used for orthodontic bonding and the group that used the previously employed acid-etch (AE) method. We extracted the frequency of brackets bonded with SEP and the frequency of adhesive failures (bond failure), as well as the frequency of brackets bonded with the AE method and the frequency of adhesive failures from each paper. We then calculated the effect size of the odds ratio (OR) using the formula below.

$$ES_{OR} = \frac{ad}{bc}$$

The concept of effect size is designed to represent different research results in a common metric so that they can be meaningfully compared when quantitatively integrating them.

C. Conduct a homogeneity test of the effect sizes of each study. Heterogeneity refers to the statistical differences in effect sizes among individual studies obtained from a systematic review. Causes of heterogeneity may include inappropriate selection of effect sizes or the presence of one or two extreme studies. After determining the effect sizes based on the type of data and conducting the analysis, the statistical heterogeneity of the results obtained should be examined. A typical method for this is to visually assess heterogeneity based on the degree of overlap of the confidence intervals of the individual studies included in the forest plot (a lack of overlap indicates the presence of heterogeneity), and to use the Q-test as a statistical test. The null hypothesis for this test is that all effect sizes among the individual studies are the same, while the alternative hypothesis is that at least one effect size is different. Considering the low statistical power due to the small number of individual studies included, a statistical significance level of 0.1 is used instead of the commonly used 0.05, and the I² statistic (Higgin., 2003) is used as a measure of heterogeneity. This is expressed as a percentage of the variation in the effect estimates, with 0-25% typically interpreted as low heterogeneity, 25-75% as moderate heterogeneity, and over 75% as substantial heterogeneity[9]. If the issue of heterogeneity is not significant, a meta-analysis can be performed; if heterogeneity is confirmed, the causes must be identified. Methods for exploring the causes of heterogeneity include subgroup analysis, meta-regression analysis, and sensitivity analysis. In the case of subgroup analysis, factors that may influence the effect estimates should be selected in advance to minimize the risk of bias in the study. If the causes of heterogeneity cannot be explained, a meta-analysis is generally not conducted, and a qualitative review is performed instead.

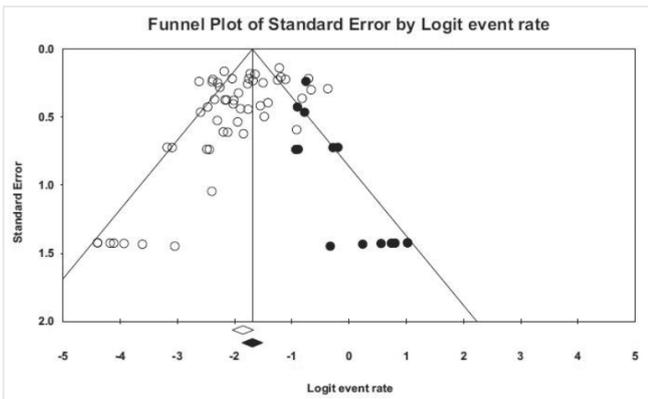

Fig. 5. Funnel plot of the 52 original studies included in the analysis(white dots) and the missing studies imputed by the trim-and-fill procedure (black dots).

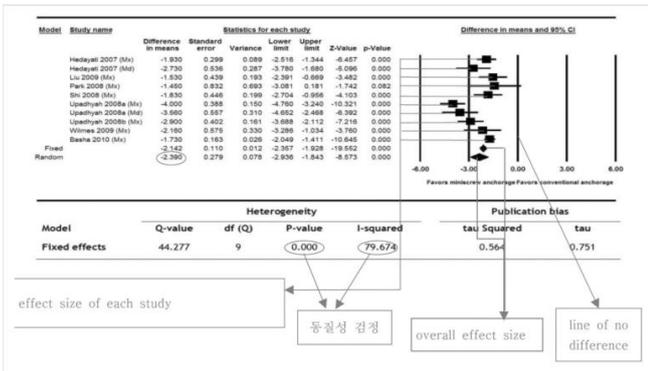

Fig. 6. Forest plot for the mean difference of the anchorage loss between the MI and conventional anchorage groups, including the number of source studies, the effect sizes with the 95% confidence intervals, the assessment of heterogeneity, and the statistical significance

D. In meta-analysis, along with estimating effect sizes, the impact of publication bias can be assessed through a funnel plot, and sensitivity analysis is conducted to examine the reliability of research findings by exploring various influences on the results. Generally, the quality assessment results are integrated to review how the quality of the research affects the research outcomes.

According to "Fig. 5", when a funnel plot for publication bias was drawn from the original studies, the white dots were asymmetrically arranged, indicating the presence of publication bias. In this case, to eliminate publication bias, the trim-and-fill procedure can be used to fill in black dots to achieve symmetry and reanalyze the data. In other words, the trim-and-fill procedure refers to adding studies that are expected to have been omitted due to publication bias, creating a left-right symmetry, and then recalculating the effect size (adjusted). If the difference between the original effect size and the adjusted effect size is small enough to be considered not significant, it provides greater confidence that the original effect size is correct.

E. *Conducting a meta-analysis and presenting the results*

Looking at the results of the meta-analysis from the dental example[10], the mean difference in anchorage loss between the two groups was -2.4mm (95% CI = (-2.9, -1.8), p = 0.00). The negative sign indicates that the mean of the Conventional Anchorage Group (CA group) was higher than that of the Miniscrew Implants Group (MI group). In other words, the anchorage loss in the MI group was lower than that in the CA group.

The figure is called a forest plot, where the size of each square corresponding to each study is determined by relative weight or sample size, and the horizontal line next to the square represents the 95% confidence interval. The center of the diamond shown in the figure represents the overall effect size, which combines the effect sizes of each study, and the horizontal line of the diamond indicates the 95% confidence interval of the overall effect size. The inclusion of 0 indicates that there is no significant difference; since 0 is not included here, it means there is a significant difference in the mean anchorage loss between the two groups "Fig. 6".

In continuous outcome variables, 0 is used to indicate no difference, while in binary outcome variables, 1 is used to indicate no effect.

Looking at the results of the homogeneity test in this example, the p-value corresponding to the Q test was 0.000, which is very significant. Additionally, the $I^2$ statistic was 79.67%, indicating the presence of heterogeneity, as it exceeds 50%. In the meta-analysis results from the dental example, the overall effect size was 1.35 (95% CI: 0.99-1.83), which did not find a significant difference in failure rates between the two groups. However, the meta-analysis comparing the time required for bonding in the SEP group and the AE group showed that the AE group took significantly 23 times longer than the SEP group. However, in 'Fig. 8", since the meta-analysis was conducted with only two studies, it did not provide sufficient evidence that time was saved in the SEP group.

Looking at the results of the homogeneity test, the $I^2$ value in "Fig. 7" was 0%, indicating no heterogeneity, while in "Fig. 8", the $I^2$ value was 68.2%, indicating the presence of heterogeneity.

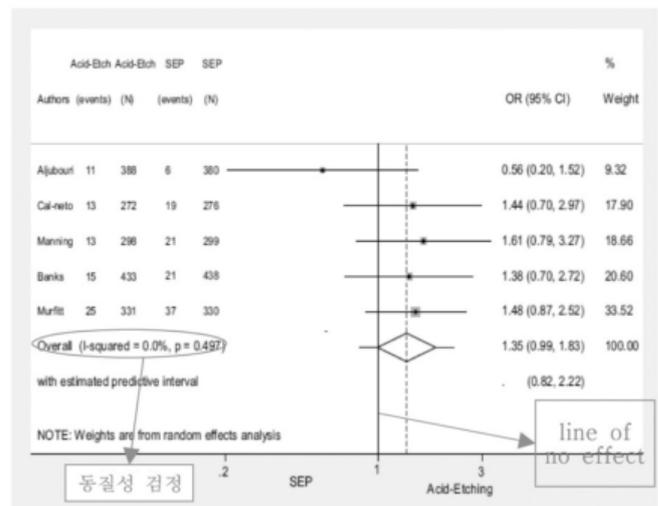

Fig. 7. Random-effects meta-analysis of bracket failure with SEP and AE4' N = total number in each group

Events = number of participants with the outcome in each group Weight = influence of studies on overall meta-analysis

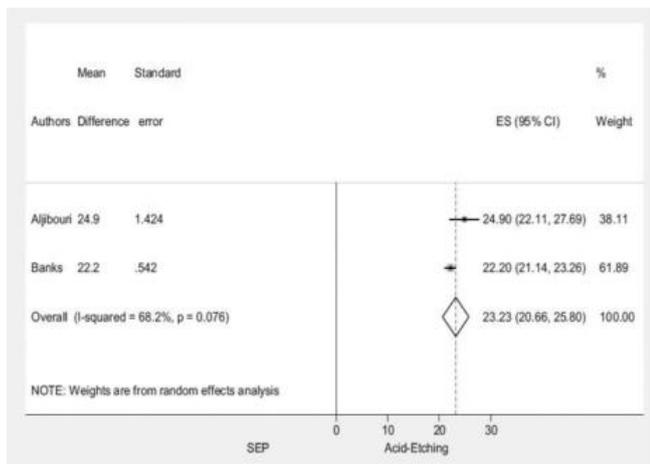

Fig. 8. Random-effects meta-analysis of required time to bond with SEP and AE

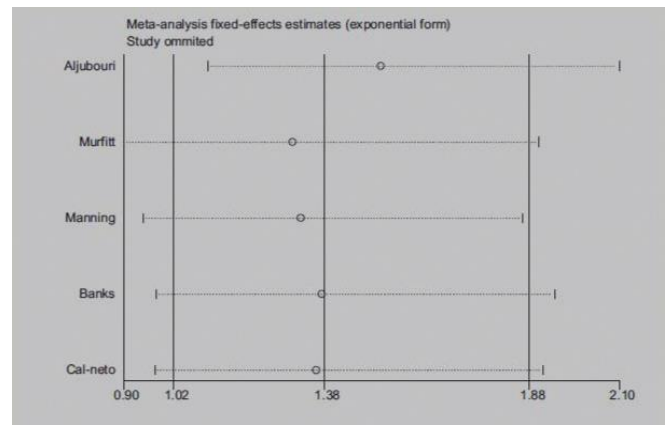

Fig. 9. Meta-analysis to investigate the influence of individual studies on the overall meta-analysis estimate

*F. Conducting Sensitivity Analysis*

Sensitivity analysis is a method used to verify whether the research results obtained from a meta-analysis are accurate. If the meta-analysis results do not change after removing one individual study from the included studies, then the results can be considered very robust. However, if the overall direction of the research results changes, it indicates that the overall trend of multiple papers is not being represented by that one study, suggesting a lack of stability in the results. In this case, it is necessary to examine whether there is any unnoticed heterogeneity in the relevant paper. The graph in "Fig. 9" is a sensitivity analysis of the results shown in "Fig. 7"

Specifically, the first line represents the results of a meta-analysis conducted with the four papers excluding the first paper by Aljubouri et al. from the five papers. The second line represents the results of a meta-analysis conducted with the four papers excluding the paper by Murfitt et al. The third, fourth, and fifth lines similarly depict the meta-analysis results excluding the respective author papers on the y-axis. Here, only the first line reports a significantly lower failure rate in AE compared to the original results. This is believed to be because the paper by Aljubouri et al. reported a lower failure rate in SEP[4].

## IV. CONCLUSION

Meta-analysis research is a method that evaluates and analyzes studies accumulated over several years, and it is a research technique that is urgently needed at this time and is highly valuable. I hope that the explanations of these procedures in various fields of dentistry will be of some help to researchers who wish to conduct meta-analysis.